\documentclass[sigconf]{acmart}
\AtBeginDocument{%
  }

\setcopyright{acmlicensed}
\copyrightyear{2018}
\acmYear{2018}
\acmDOI{XXXXXXX.XXXXXXX}
\acmConference[Conference acronym 'XX]{Make sure to enter the correct
  conference title from your rights confirmation email}{June 03--05,
  2018}{Woodstock, NY}
\acmISBN{978-1-4503-XXXX-X/2018/06}

\usepackage{booktabs}
\usepackage{tabularx}
\usepackage{makecell}

\usepackage[most]{tcolorbox} 
\definecolor{block-gray}{gray}{0.96}

\newtcolorbox{sprompt}[2][]{%
    colback=block-gray,
    boxrule=0pt,
    boxsep=0pt,
    breakable,
    enhanced jigsaw,
    borderline west={4pt}{0pt}{gray},
    title={#2\par},
    colbacktitle={block-gray},
    coltitle={black},
    fonttitle={\large\bfseries},
    attach title to upper={},
    #1,
}

\definecolor{block-blue}{HTML}{EBF7FF}
\definecolor{accent-blue}{HTML}{569FD2}
\newtcolorbox{hprompt}[2][]{%
    colback=block-blue,
    boxrule=0pt,
    boxsep=0pt,
    breakable,
    enhanced jigsaw,
    borderline west={4pt}{0pt}{accent-blue},
    title={#2\par},
    colbacktitle={block-blue},
    coltitle={black},
    fonttitle={\large\bfseries},
    attach title to upper={},
    #1,
}

\begin{document}
\title[GhostWriter: Augmenting Human-AI Writing Experiences Through Personalization and Agency]{GhostWriter: Augmenting Collaborative Human-AI Writing Experiences Through Personalization and Agency}

\author{Catherine Yeh}
\email{catherineyeh@g.harvard.edu}
\affiliation{%
  \institution{Harvard University}
  \city{Cambridge}
  \state{Massachusetts}
  \country{USA}
}

\author{Gonzalo Ramos}
\affiliation{%
  \institution{Microsoft Research}
  \city{Redmond}
  \state{Washington}
  \country{USA}}
\email{goramos@microsoft.com}

\author{Rachel Ng}
\affiliation{%
  \institution{Microsoft}
   \city{Redmond}
  \state{Washington}
  \country{USA}}
\email{rng@microsoft.com}

\author{Andy Huntington}
\affiliation{%
  \institution{Microsoft}
   \city{London}
  \country{UK}}
\email{anhuntin@microsoft.com}

\author{Richard Banks}
\affiliation{%
  \institution{Microsoft Research}
   \city{Cambridge}
  \country{UK}}
\email{rbanks@microsoft.com}

\begin{abstract}
Large language models (LLMs) have become ubiquitous in providing different forms of writing assistance to different writers. 
However, LLM-powered writing systems often fall short in capturing the nuanced personalization and control needed to effectively support users -- particularly for those who lack experience with prompt engineering.
To address these challenges, we introduce GhostWriter, an AI-enhanced design probe that enables users to exercise enhanced agency and personalization during writing. 
GhostWriter leverages LLMs to implicitly learn the user's intended writing style for seamless personalization, while exposing explicit teaching moments for style refinement and reflection.
We study 18 participants who use GhostWriter on two distinct writing tasks, observing that it helps users craft personalized text generations and empowers them by providing multiple ways to control the system's writing style. 
Based on this study, we present insights on how specific design choices can promote greater user agency in AI-assisted writing and discuss people's evolving relationships with such systems. 
We conclude by offering design recommendations for future work.

\end{abstract}



\begin{teaserfigure}
  \includegraphics[width=\textwidth]{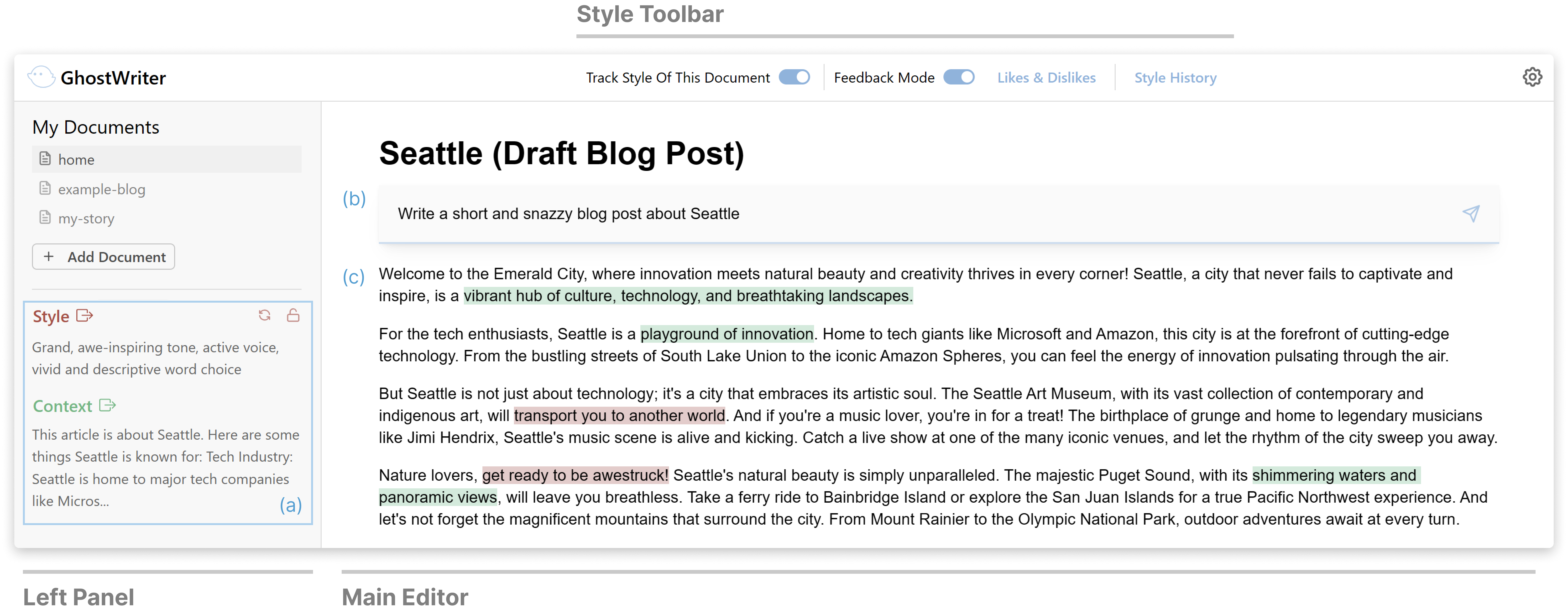}
  \caption{Overview of GhostWriter, an AI-powered environment that personalizes the writing process through \textbf{(a)} editable \textit{style} and \textit{context} information. \textbf{(b)} Personalized content can be generated using features such as inline LLM prompts. \textbf{(c)} Users can also explicitly teach the system about their style preferences by highlighting likes \& dislikes.}
  \Description{Overview of GhostWriter, an AI-powered environment that personalizes the writing process. The left panel contains a list of documents and the user's editable style and context information. The main editor is in the center of the interface, where users can write text and use features such as inline LLM prompts. There is also a style toolbar at the top of the interface with additional customization options.}
  \label{fig:teaser}
\end{teaserfigure}

\received{20 February 2007}
\received[revised]{12 March 2009}
\received[accepted]{5 June 2009}

\maketitle

\section{Introduction}\label{sec:intro}
With the rise of Large Language Models (LLMs), there is growing interest in leveraging their text generation capabilities~\cite{brown2020language,lee2024design} to enhance the process and product of writing. 
Many emerging systems use LLMs to assist with various writing tasks, from brainstorming and drafting content~\cite{yuan2022wordcraft,mirowski2023co,kim2023metaphorian,wan2024felt} to summarizing~\cite{dang2022beyond,zhang2023benchmarking,zhang2023summit} and refining existing text~\cite{raheja2023coedit,yuan2022wordcraft}.

These systems have immense potential to augment writing experiences, but also introduce persistent challenges.
To illustrate, imagine Sarah, a software engineer looking to use AI to help write a short introduction email to her new team at work.
When asking \href{https://chat.openai.com/}{ChatGPT} for assistance, she might receive the following as output:

\begin{sprompt}{}
\small
\textbf{Subject: Introduction and Excitement for Joining the Team!}
\vspace{0.25em}

     Dear Team,
     
\vspace{0.25em}

    I hope this email finds you all in good spirits. My name is Sarah, and I am thrilled to introduce myself as the newest member of the team. I am joining as a software engineer, and I couldn't be more excited to embark on this journey with all of you...
    \vspace{0.25em}
\end{sprompt}

This example illustrates two key challenges.
First, the output can be too general, reflecting a \textbf{lack of personalization} in the generated text~\cite{yuan2022wordcraft,mirowski2023co,kim2023repurposing,gero2023social,chung2022talebrush,wasi2024ink,anderson2024homogenization,mysore2023pearl}. 
The LLM likely did not write in Sarah's voice, making the output seem ``average'' and as if it came from someone else~\cite{padmakumar2023does}.
Additionally, the LLM may be missing relevant context about Sarah or her situation that may be helpful to know when writing this email
(e.g., her knowledge about a topic, what project she is working on with the new team, etc.). 

Second, the complexity of articulating a prompt can elicit feelings of \textbf{limited agency}\footnote{Interpreted as the action of steering a system behavior toward a desired outcome with confidence -- distinct from the more low-level notion of control.} when working with LLMs~\cite{yuan2022wordcraft,chung2022talebrush,ippolito2022creative,mahlow2023writing,clark2018creative,zamfirescu2023johnny}. 
If Sarah, for example, wants to change the writing style of the generated email, it can be difficult for her to make those changes, as this requires one to understand what needs to change and why~\cite{yuan2022wordcraft}. 
An LLM's probabilistic nature can also make Sarah feel uncertain about how her edits will influence the model's behavior~\cite{lee2022coauthor,lu2021fantastically}, as minor prompt alternations can lead to major changes in output~\cite{salinas2024butterfly}.

These challenges of personalization and agency in the use of LLMs can limit their benefits when applied to writing and underscore opportunities for more meaningful co-creation with these systems.
We argue that understanding and addressing these challenges requires not only technical advances, but also tools that provoke reflection on how users interact with AI during writing~\cite{morris2023design}.
We see design and technology probes \cite{gaver1999design, boehner2007hci, graham2008probes, hutchinson2003technology} -- exploratory tools intended to elicit insights into user values, practices, and interactions with technology -- as key instruments to meet, understand, and ultimately address these challenges.

In this work, we introduce \textbf{GhostWriter}, a \textit{design probe} that explores how AI writing tools can support user agency and personalization. Specifically, GhostWriter aims to help us investigate new human-AI interaction ideas around these design goals:
\begin{itemize}
    \item \textbf{DG1:} Provide rich personalization options and feedback to align LLM-based writing outcomes with user intents.
    \item \textbf{DG2:} Expose and champion user agency in AI-powered writing interfaces.
\end{itemize}

Embodied as an AI-powered editor, GhostWriter gives people an LLM-augmented writing experience by providing agency in \textit{style} and \textit{context} personalization through implicit and explicit means (\autoref{fig:teaser}). 
Having the ability to steer the model toward an intended style can help users like Sarah generate text that is more aligned with their goals and intentions, rather than relying on the LLM's default ``generic'' style.
Communicating in a personalized style is central to many writing tasks~\cite{yuan2022wordcraft,clark2018creative}, serving as an expression of one's voice, identity, and rhetorical goals.
We believe one way to provide agency over style is through contextual guidance: by shaping the context given to the model, users can more effectively influence the style and content of its output.

Following methods from works such as~\citet{hohman2019gamut}, we use GhostWriter as a probe to evaluate our design ideas to support a personalized experience that empowers users with the ability to shape LLM outputs and infuse their writing with a desired style while co-editing with AI. 
In a user study, we examined how 18 participants reacted to and used GhostWriter in two tasks that represent a good cross-section of writing as an activity: professional editing and creative writing. 

Our results reveal that GhostWriter helped users exert control over the direction of LLM outputs and offered value in providing them with flexible ways to customize style and context. 
We observed that different methods of style control were useful at different points in a task, and across different kinds of writing.
At the same time, our study surfaced challenges around interpreting the system's behavior during style personalization and navigating questions about outcome ownership when writing with AI. 
These findings informed a set of design recommendations for creating AI-infused tools that amplify human intent while preserving ownership.

Our contributions include:
\begin{itemize}
    \item \textbf{GhostWriter}, an AI-powered writing experience that supports content personalization through style and context specifications.
    \item \textbf{User study results} from using GhostWriter as a probe that indicate the perceived utility and effectiveness of our system in generating personalized text and championing user agency.
    \item \textbf{Design recommendations} and directions to help shape the design of future LLM-infused writing systems.
\end{itemize}

\section{Related Work}\label{sec:related}
\subsection{AI-Assisted Writing}
Much work investigates how AI can enhance human writing~\cite{kim2023repurposing,dang2022beyond,yang2022ai,fang2023systematic,chung2022talebrush,gero2023social,kim2023metaphorian,hoque2023portrayal,wang2024wordflow,stark2023can} across various users and tasks~\cite{lee2024design}. 
\citet{clark2018creative} explores how a machine-in-the-loop system can amplify creativity for short story and slogan writing tasks, and \citet{singh2022hide} present a multimodal interface for creative writing powered by generative AI. Wordcraft~\cite{yuan2022wordcraft,ippolito2022creative} is an LLM-augmented writing editor, offering features such as rephrasing or continuing a text passage, and just-in-time custom controls. 
Dramatron~\cite{mirowski2023co} also uses LLMs for creative text generation, exploring the use of these models to co-write screenplays.
However, a common theme in these systems is the lack of context awareness and user agency when generating text, which we aim to address. 

System extensions, such as Gmail Smart Compose~\cite{chen2019gmail}, Grammarly,\footnote{\url{https://www.grammarly.com/ai-writing-tools}} and Wordtune~\cite{zhao2022leveraging}, integrate AI features into existing experiences to support writing suggestions and corrections.
Other platforms like \href{https://notion.so}{Notion} have introduced document search, text generation, and style analysis features, which capture a vision similar to ours for AI-assisted writing. 
Our work with GhostWriter, developed in parallel, relates to these design patterns but takes a different path, focusing more on style capture, context definition, and output personalization.

\subsection{Personalized AI Experiences}
\subsubsection{Explicit vs. Implicit Personalization}
Personalization can be achieved through implicit or explicit methods. 
Many recommendation systems use \textit{implicit} forms of personalization to model user preferences over time, by looking at user interaction histories~\cite{li2023gpt4rec}, assessment performance~\cite{benhamdi2017personalized}, or mouse click patterns~\cite{kim2011recommender}.
Some AI systems also incorporate \textit{explicit} approaches such as having users bookmark~\cite{rachatasumrit2021forsense,park2023foundwright} or react to content~\cite{narayanan2023understanding} to receive personalized suggestions while retaining user agency.

We take inspiration from and aim to combine both kinds of personalization in having GhostWriter implicitly learn the user's style as they write, while also providing opportunities for explicit style refinement.
Extending the concept of ``natural language user profiles'' from recommendation systems~\cite{radlinski2022natural,mysore2023editable}, we explore creating editable style and context profiles to guide the generation of personalized text. 
While Impressona~\cite{benharrak2023writerdefined} creates personas with different writing styles to generate stylized \textit{feedback}, we focus on personalizing the written \textit{content} itself -- a core challenge and need when writing with LLMs~\cite{mysore2023pearl,padmakumar2023does}.

\subsubsection{Personalization Without Retraining}
Our work builds on the paradigm of personalizing downstream user experiences without retraining the underlying ML models~\cite{park2023foundwright,rachatasumrit2021forsense}. 
This approach is powerful because it allows for flexible, user-specific customization without compromising the performance or scalability of pretrained models.
In writing contexts, recent research explores how LLMs can help generate personalized text without model retraining~\cite{salemi2023lamp,zhong2022less,kim2024few}. 
Some work proposes incorporating social factor modeling to personalize the writing experience with LLMs~\cite{kulkarni2023writing}, while others draw on principles from writing education~\cite{li2023teach}. 
LMCanvas~\cite{kim2023lmcanvas} transforms the traditional text editor into a canvas-based interface, where malleable ``blocks'' are used to create a personalized writing environment. 
We offer a complementary perspective by studying personalization through the lens of writing style and context.

We also draw on ideas from interactive machine teaching~\cite{ramos2020interactive, simard2017machine},
where a human teacher communicates information to a machine learner in an iterative process that has been shown to enhance both the user experience and the building of an efficient learning set~\cite{taneja2022framework,zhou2022exploiting}.
With GhostWriter, users iteratively and interactively construct their target style and context to guide the generation of (personalized) text.

\subsection{Working with Style and Context}
GhostWriter can infer a writing style from a written sample using an LLM -- a human-interpretable alternative to neural methods such as style representation learning~\cite{patel2023learning} and activation steering~\cite{konen2024style}. 
Another relevant idea from NLP is text style transfer (TST)~\cite{jin2022deep,hu2022text,fu2018style}, which aims to preserve the content of generated text while adjusting attributes like tone or voice. While TSTs are limited to preexisting text, our system can apply learned styles to produce new text, similar to~\citet{li2023teach}. 
A related task is controllable text generation (CTG), where various aspects of generated text can be manipulated, such as context~\cite{sordoni2015neural} or topic~\cite{dziri2018augmenting}. 
Our work explores how natural language can enable end-users to ``perform'' personalized TSTs and controllable text generation with LLMs. 
We also draw from context-faithful prompting~\cite{zhou2023context} when incorporating the user's writing style and context.

ChatGPT's custom instructions feature enables users to specify information that the system should consider when generating output.\footnote{\url{https://openai.com/blog/custom-instructions-for-chatgpt}. Announced on July 20, 2023, post our original GhostWriter implementation.}
Similarly, we allow users to tweak context and style information through natural language, but our system uses these details to generate personalized \textit{writing} rather than \textit{chat output}. 
While some of these features may overlap, the user experiences (i.e., separated chat vs. in-document interactions) remain distinct, and we provide tangible insights about their perceived value through our user studies.

\section{System Overview}\label{sec:system}
We address \textbf{personalization} and \textbf{agency} in the context of AI-assisted writing through GhostWriter, an LLM-powered writing environment.
We use GhostWriter as a design probe~\cite{boehner2007hci,hutchinson2003technology,graham2008probes} to explore the possibility of using AI to craft personalized outputs through the manipulation of \textit{style} and \textit{context} in ways that emphasize user agency. 

Design probes in HCI are adapted from cultural probes~\cite{gaver1999design}, designed objects that promote participant engagement in the design process~\cite{boehner2007hci,rachatasumrit2021forsense,jorke2023pearl,park2023foundwright,hohman2019gamut}. 
These probes focus on identifying areas of value and improvement, rather than evaluating usability or comparing against existing solutions~\cite{park2023foundwright}.
As such, GhostWriter helps to elicit feedback about how our design and implementation choices meet the challenges of personalization and agency. 

\begin{figure*}
    \centering
    \includegraphics[width=\linewidth]{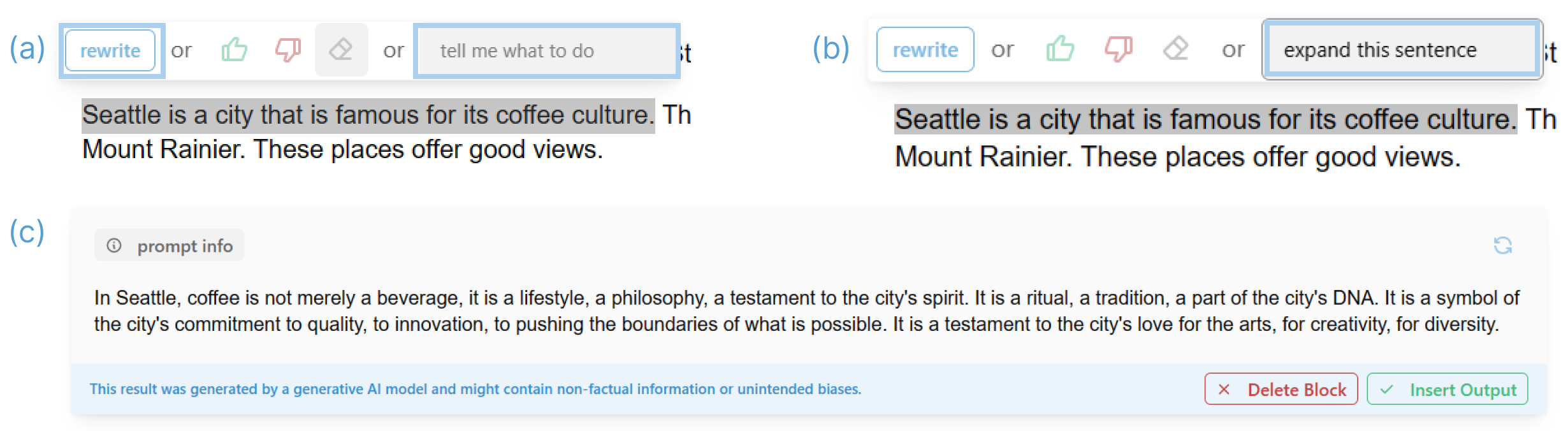}
    \caption{Personalization through refining text. \textbf{(a)} Upon invoking the context menu, the user can choose the \textit{rewrite} or the \textit{``apply'' prompt} option. \textbf{(b)} The ``apply'' prompt will apply the inputted text as a prompt to the current text selection. \textbf{(c)} Sample output from using an ``apply'' prompt. Users can regenerate, delete, or insert the outputted text.}
    \Description{In the context menu, users can rewrite the selected text, or prompt for more specific changes. We show sample output from using an ``apply'' prompt after expanding a sentence about Seattle's famous coffee culture.}
    \label{fig:context-menu}
\end{figure*}
\begin{figure*}
    \centering
    \includegraphics[width=\linewidth]{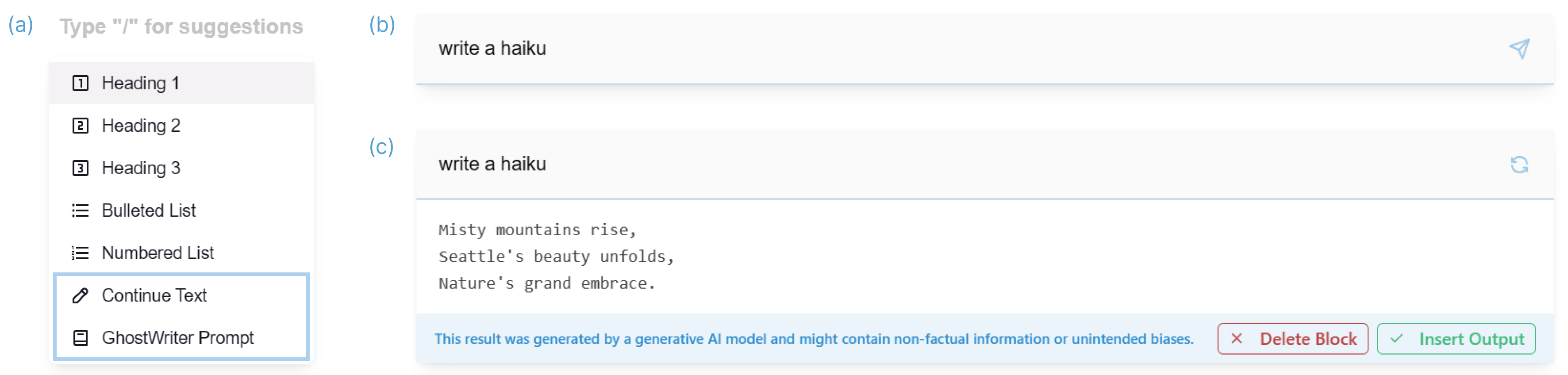}
    \caption{Personalization through generating new text. \textbf{(a)} Upon invoking the slash menu, the user can choose the \textit{continue text} or \textit{inline (``GhostWriter'') prompt} option. \textbf{(b)} The inline prompt takes any general prompt as input. \textbf{(c)} Sample output from an inline prompt. Users can regenerate, delete, or insert the outputted text.}
    \Description{With the slash menu, users can continue the current text or use an inline prompt to generate more content. We show sample output from an inline prompt after writing a haiku about Seattle.}
    \label{fig:slash}
\end{figure*}
\subsection{Design Principles}
The design, development, and study of GhostWriter was guided by our design goals (Section~\ref{sec:intro}). 
These goals led us to ideate how we could enhance agency and personalization in an LLM-powered writing experience. 
We propose that these goals can be met by considering both implicit (i.e., learning as you write) and explicit (i.e., learning from direct user input) forms of feedback, which together can help yield more goal-aligned LLM outputs through the iterative definition of context and writing style.
Building on this strategy and informed by existing literature on AI writing systems, we define the following design principles (DPs): 

\subsubsection{\textbf{DP1:} Leverage Machine Capabilities While Championing Agency}
Our work explores opportunities to use LLMs' text generation and analysis capabilities~\cite{brown2020language,yuan2022wordcraft,mirowski2023co} to extract writing styles that can be used to produce bespoke outputs \textbf{[DG1]}.
At the same time, we aim to champion user agency when co-writing with AI \textbf{[DG2]}. 
This position places technology in the service of human expression and can foster a writing experience that addresses previous concerns about personalization and control~\cite{yuan2022wordcraft,chung2022talebrush,ippolito2022creative}.

\subsubsection{\textbf{DP2:} Use Familiar Editor Metaphors}
To reduce the cognitive load~\cite{sweller2011cognitive} of navigating a new system, 
we use existing text editor metaphors in our design~\cite{nielsen2005ten} (e.g., highlighting text, section blocks).
By exploring how AI augmentations blend into familiar writing experiences, we can focus on distilling their effects and project our learnings into grounded, relatable scenarios \textbf{[DG1]}.  

\subsubsection{\textbf{DP3:} Blend Into the Writer's Existing Workflow}
Users should not have to deviate from existing flows when using our system. 
We prioritize simplicity and a non-fragmented user experience, like Notion or \href{https://www.microsoft.com/en-us/microsoft-loop}{Microsoft Loop}, which offer straightforward, yet feature-rich writing interfaces \textbf{[DG1]}. 
In these systems, writing and most interactions occur in one central editor, reducing interface complexity and unnecessary context-switching~\cite{mark2008cost}.

\subsubsection{\textbf{DP4:} Provide Transparency to Support Reflection and Discovery}
GhostWriter strives to offer transparency about its internal state. 
Users should know and be able to inspect what information the LLM has access to, which can be achieved in an easy-to-understand way through natural language~\cite{radlinski2022natural,mysore2023editable}.
In addition to fostering reflection and experimentation with alternative styles and contexts~\cite{mahlow2023writing}, this transparency can help users understand what they can do with the system and how to fix problems when they arise -- a key aspect of championing agency~\cite{amershi2019guidelines} \textbf{[DG2]}.

\subsection{Interface Design}
We outline the key components of GhostWriter's interface, as shaped by our DPs. 
Our design and interaction strategies are broadly applicable to LLMs, and present an alternative to emerging experiences that rely on linear, turn-based chat interfaces.

\begin{figure*}
    \centering
    \includegraphics[width=\linewidth]{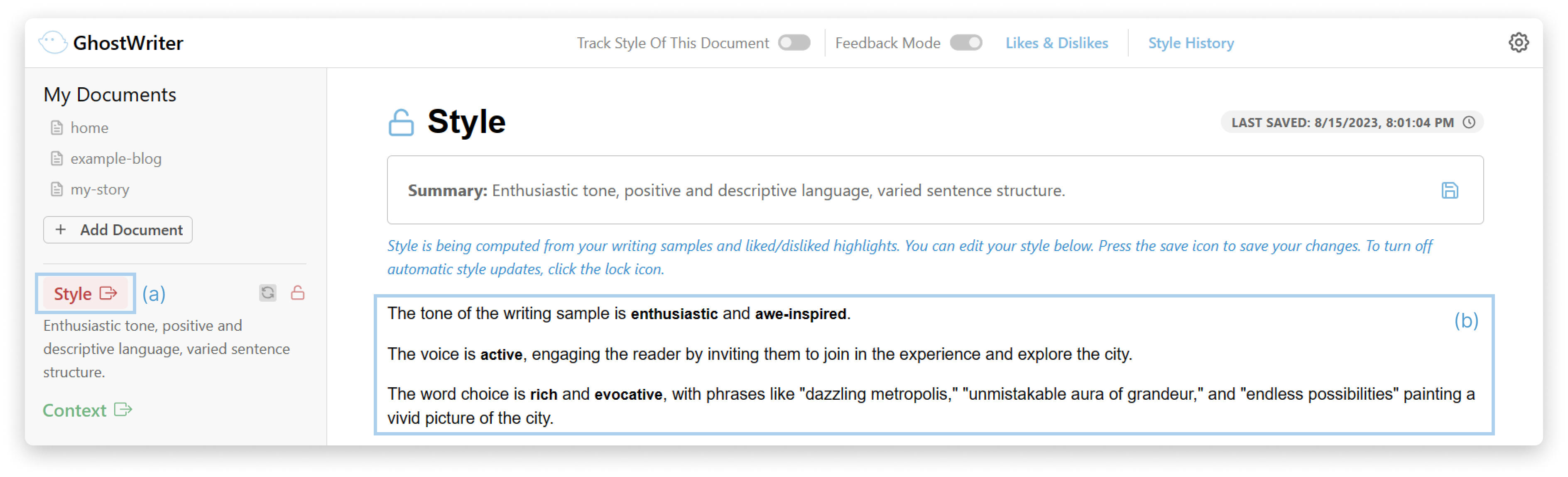}
    \caption{Teaching style through manual edits. \textbf{(a)} Users can view and edit the full description of their current style by pressing the Style button. \textbf{(b)} The style description (excerpt shown) is editable like a normal text file.}
    \Description{Interface screenshot showing the full description of the user's current style. The style description (excerpt shown) is editable like a normal text file.}
    \label{fig:edit-style}
\end{figure*}

\subsubsection{Main Editor}
GhostWriter's central view is the \textbf{main editor} (\autoref{fig:teaser}), which mirrors existing text editors \textbf{[DP2]} and provides a space for users to author documents. One way users can teach GhostWriter about their target writing style is simply by writing. After each \textit{n} (default: 100) new characters, the system analyzes the document to extract its style \textbf{[DP1]}.
Users can apply LLM-powered features to refine existing text or generate new text given the current writing style and context \textbf{[DP1]} (\autoref{tab:llm-features}). 
All features are embedded inline in the editor to support a non-fragmented, familiar writing experience \textbf{[DP2, DP3]}.

To \textit{refine} written text, users can invoke the context menu by selecting a portion of the document. Then, they can rewrite or use a contextual ``apply'' prompt to refine the selected text (\autoref{fig:context-menu}a). 
Both operations apply the current style to generate personalized content (\autoref{fig:context-menu}b). 
To \textit{generate} text, users can invoke the slash menu by typing a forward slash (``/'') anywhere in the document (\autoref{fig:slash}a). Then, they can continue the text from the current point or generate new text by invoking the inline ``GhostWriter'' prompt. 
These operations use both the current learned style \textit{and} context to generate content
(\autoref{fig:slash}b). 
For all context and slash menu features, users can regenerate, delete, or insert LLM output \textbf{[DP1]} (\autoref{fig:context-menu}c, \autoref{fig:slash}c). 

\begin{figure}
    \centering
    \includegraphics[width=\linewidth]{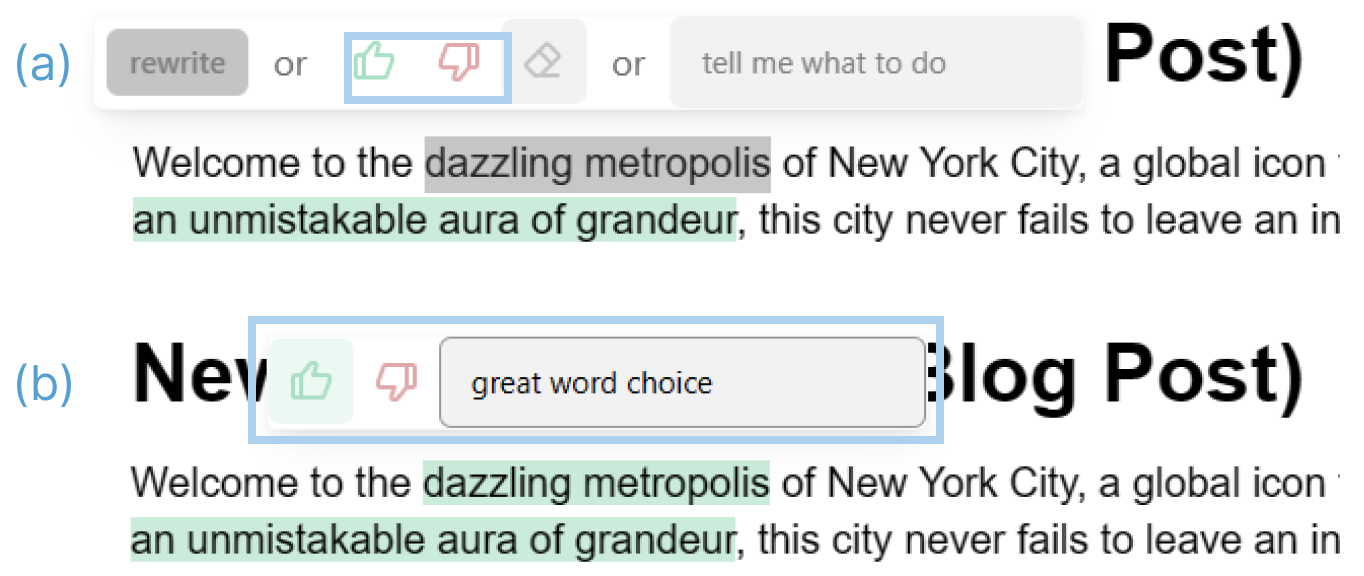}
    \caption{Teaching style through likes and dislikes. \textbf{(a)} Users can like (thumbs up) or dislike (thumbs down) any text selection through the invoked context menu. \textbf{(b)} Once the corresponding icon is selected, the user can optionally provide feedback as to why they like or dislike the highlighted text.}
    \Description{In the context menu, users can also teach style through likes and dislikes with the corresponding thumbs up or thumbs down icons. The user can optionally provide feedback as to why they like or dislike the highlighted text (e.g., ``great word choice'').}
    \label{fig:highlight}
\end{figure}

The user can explicitly teach GhostWriter about their target writing style by indicating likes and dislikes \textbf{[DP1]}. 
They do so by highlighting a portion of text \textbf{[DP2]}, which invokes the context menu (\autoref{fig:highlight}a). 
Then, they can like or dislike the selection and optionally write \textit{why} to help the system learn about their style (\autoref{fig:highlight}b).
Allowing users to explicitly nudge GhostWriter toward their preferences builds on work showing the value of fine-grained feedback for personalization~\cite{rachatasumrit2021forsense,narayanan2023understanding}. 
We also draw from active reading practices, where annotation and highlighting are key strategies in engaging with and reflecting on the text~\cite{schilit1998beyond}. 

\begin{table}
\small
\caption{GhostWriter's LLM-powered features for personalized text generation.}
\label{tab:llm-features}
\begin{tabularx}{\linewidth}{p{.26\linewidth}p{.66\linewidth}}
\textbf{Feature} & \textbf{Description} \\ \midrule
\parbox[t]{\linewidth}{\textbf{Rewrite} \\ (Context menu)}        & Rewrite text selection to match the system's learned \textit{style}                   \\ 
\parbox[t]{\linewidth}{\textbf{``Apply'' prompt} \\ (Context menu)}    & Apply LLM prompt to text selection to update its content using the system's learned \textit{style}       \\ 
\parbox[t]{\linewidth}{\textbf{Continue} \\ (Slash menu)}      & Generate new text to continue the current document using the system's learned \textit{style} and \textit{context}       \\ 
\parbox[t]{\linewidth}{\textbf{Inline prompt} \\ (Slash menu)}      & Generate new content based on an LLM prompt using the system's learned \textit{style} and \textit{context}      \\ 
\end{tabularx}
\end{table}
\begin{figure*}
    \centering
    \includegraphics[width=\linewidth]{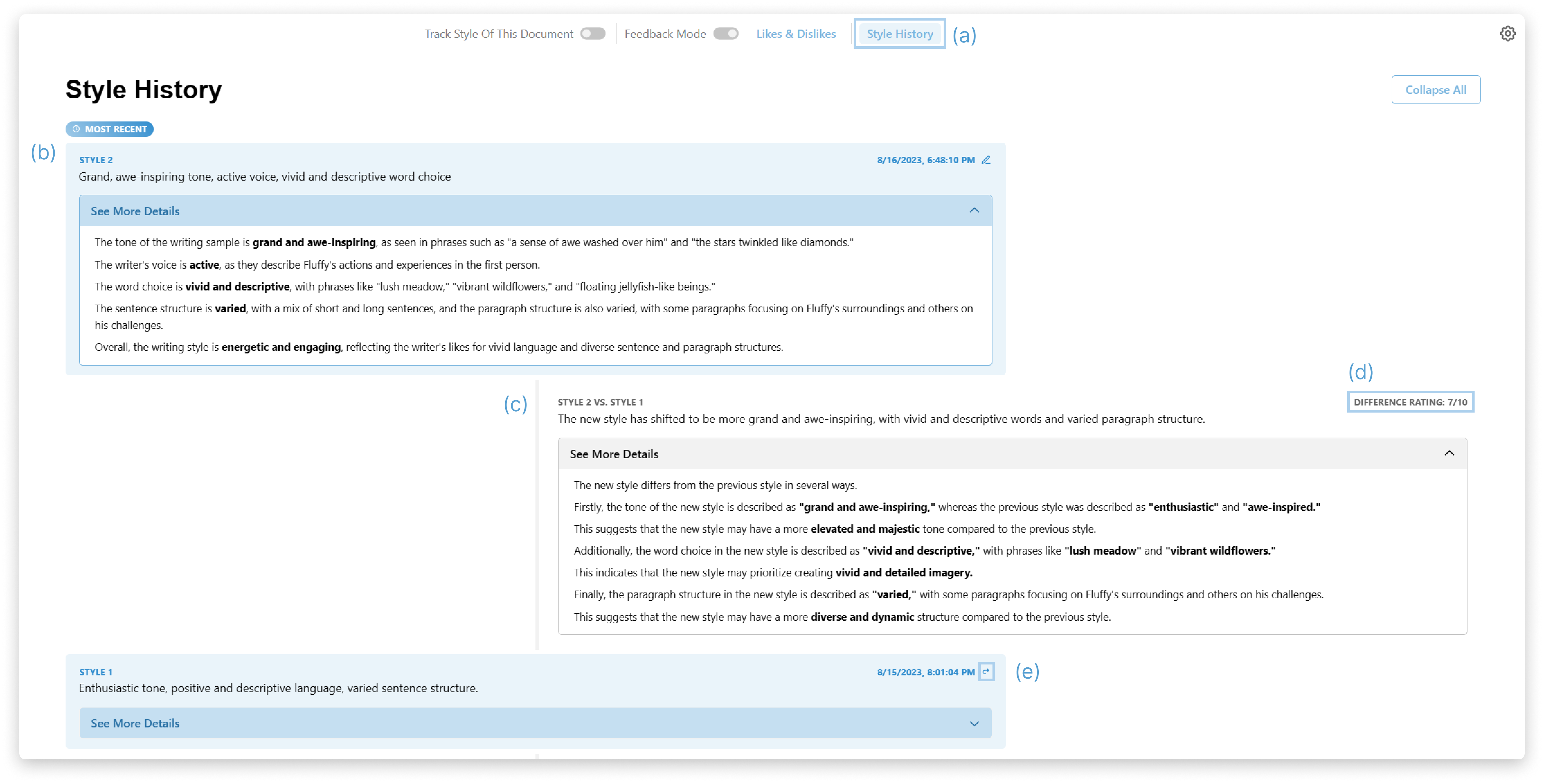}
    \caption{\textbf{(a)} Users can view all their past writing styles on the Style History page. \textbf{(b)} The blue boxes on the left each display a writing style and \textbf{(c)} the gray boxes on the right display a comparison of each pair of adjacent writing styles. \textbf{(d)} Each comparison also includes the difference rating between adjacent styles. \textbf{(e)} Users can revert to prior writing styles as well.}
    \Description{Screenshot of the Style History page, where users can view all their past writing styles. The blue boxes on the left each display a writing style and the gray boxes on the right display a comparison of each pair of adjacent writing styles.}
    \label{fig:history}
\end{figure*}

\subsubsection{Left Panel}
In the \textbf{left panel}, users can view their \textit{document} list \textbf{[DP2, DP3]} and explore the system's current \textit{style and context} \textbf{[DP4]} (\autoref{fig:teaser}). 
The style summary is automatically refreshed when the system's writing style updates.
Clicking the refresh icon will force a style update based on the current document \textbf{[DP1]}. 
Users can disable automatic style updates by pressing the lock icon.

GhostWriter avoids a cold start by providing a default (generic) writing style that evolves based on user input and interaction. When extracting style from a sample, the system is prompted to analyze five style characteristics: \textit{tone}, \textit{voice}, \textit{word choice}, \textit{sentence structure}, and \textit{paragraph structure}. We chose these dimensions to establish a relatable language for communicating style, based on informal pilot studies and work such as~\citet{reinhart2025llms}.

Users can edit the system's style directly \textbf{[DP1]} by pressing ``Style'' in the left sidebar (\autoref{fig:edit-style}a). This opens the full style description in the main editor \textbf{[DP3, DP4]} (\autoref{fig:edit-style}b). 
In allowing users to inspect and modify the system's style, we aim to support reflection and experimentation during writing \textbf{[DP4]}.
To (optionally) provide additional context for grounding text generations, users can edit the ``Context'' page in the main editor \textbf{[DP1, DP3, DP4]} (\autoref{fig:teaser}). There are no constraints on what can be included as ``context,'' and the user has full agency over this page \textbf{[DP1]}.

\begin{figure*}
    \centering
    \includegraphics[width=\linewidth]{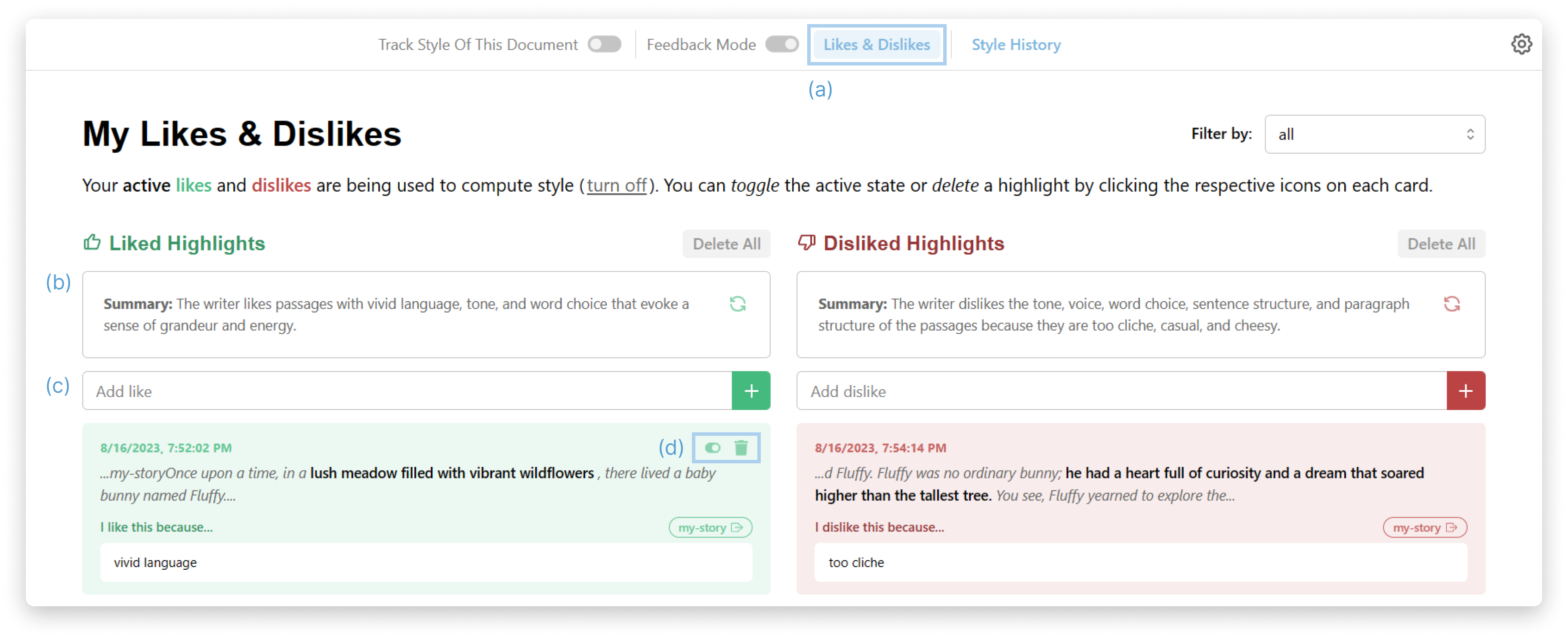}
    \caption{\textbf{(a)} Users can view all their likes and dislikes on the Likes \& Dislikes page. \textbf{(b)} A system-generated summary of the user's likes and dislikes is displayed at the top of the page. \textbf{(c)} The user can also manually add additional likes and dislikes. \textbf{(d)} Upon hovering on a like or dislike card, users have the option of toggling its active state or deleting it from their collection.}
    \Description{Screenshot of the Likes \& Dislikes page. A system-generated summary of the user's likes (highlighted in green) and dislikes (highlighted in red) is displayed at the top of the page.}
    \label{fig:likes}
\end{figure*}
\begin{figure*}
    \centering
    \includegraphics[width=0.9\linewidth]{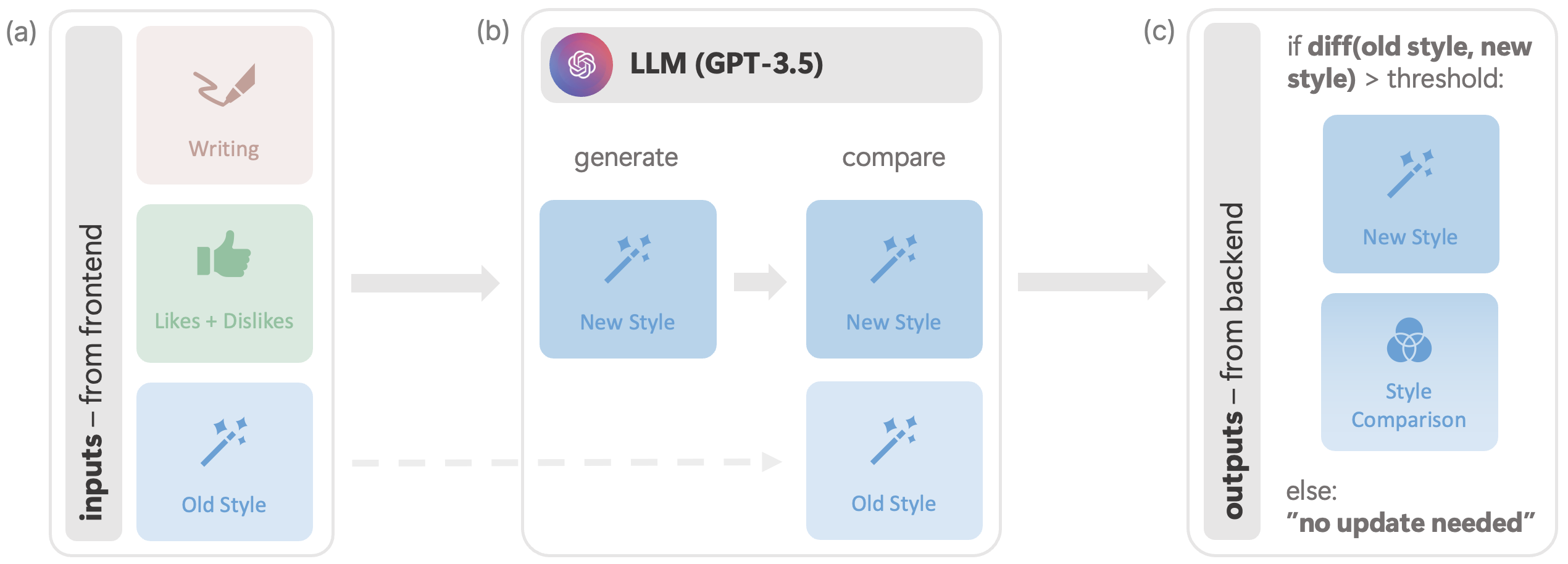}
    \caption{How style updates are computed by GhostWriter. \textbf{(a)} The current document, likes \& dislikes, and style description are passed as inputs from the frontend. \textbf{(b)} In the backend, we then ask the LLM to generate a new style description given this information. The LLM also generates a style comparison between the old and new styles, and computes a difference rating. \textbf{(c)} If the difference rating is greater than some threshold (e.g., 3 out of 10), the new style and comparison are passed as outputs back to the frontend. 
    Otherwise, the user will be informed that there is no style update needed.
    }
    \Description{Overview of how style updates are computed by GhostWriter. The current document, likes \& dislikes, and style description are passed as inputs from the frontend. In the backend, the LLM generates a new style description and comparison with this information.}
    \label{fig:style-update}
\end{figure*}

\subsubsection{Style Toolbar}
The \textbf{style toolbar}, located above the main editor, exposes features to customize the user's experience \textbf{[DP1]} and inspect the system's style knowledge \textbf{[DP4]} (\autoref{fig:teaser}).
Users can toggle the ``Track Style Of This Document'' flag to turn on/off automatic style updates for the \textit{current} document (vs. the \textit{global} style lock in the left panel). They can also turn ``Feedback Mode'' on/off, which shows or hides all highlights in the current document. 

Users also have the option to examine current and past writing styles through the ``Style History'' page (\autoref{fig:history}a) \textbf{[DP3]}, where more recent styles are displayed at the top. 
\textit{Styles} are displayed in blue boxes on the left (\autoref{fig:history}b), while \textit{comparisons} between each pair of adjacent styles are shown in gray boxes on the right (\autoref{fig:history}c). 
These LLM-generated comparisons \textbf{[DP1]} offer deeper insight into how the system-learned style changed over time by providing a ``difference rating'' (\autoref{fig:history}d) that quantifies the difference between adjacent styles from 0 to 10 (0: identical, 10: entirely different) \textbf{[DP4]}. 
Users can revert to a previous style by clicking the \textit{revert} icon in any style box (\autoref{fig:history}e) \textbf{[DP1]}.

Similarly, users can view their collection of highlighted ``Likes \& Dislikes'' (\autoref{fig:likes}a) \textbf{[DP3]}. 
At the top of this page, there are LLM-generated summaries of the user's (qualified) liked and disliked text (\autoref{fig:likes}b) \textbf{[DP1]}. 
These summaries can encourage additional reflection \textbf{[DP4]} and help users assess whether GhostWriter correctly understands their feedback.
Users can add additional likes and dislikes (\autoref{fig:likes}c) or toggle/delete highlights in this view (\autoref{fig:likes}d) \textbf{[DP1]}. 
Only active highlights are used to compute the like/dislike summaries, which in turn guide text generations. 

\subsection{Implementation}
GhostWriter is a React web application connected to a Python backend through RESTful endpoints. The main editor interface is built using \href{https://tiptap.dev/}{Tiptap}, a headless editor framework.

\subsubsection{Backend AI Services}
We orchestrated all backend LLM operations with \href{https://langchain.com/}{LangChain}, most of which use \texttt{GPT-4}. 
Style updates are computed using \texttt{GPT-3.5-Turbo},\footnote{Both were among the leading models from OpenAI at the time of implementation.} which provided comparable, but faster results during experimentation. 

We iteratively crafted and refined the prompts used by GhostWriter (included as Supplementary Material). 
Given our goal of producing a design probe, we did not optimize them to achieve perfect outcomes.
Instead, we used \texttt{GPT-4} to evaluate the generated output as a sanity check to supplement qualitative inspections~\cite{zheng2023judging}. 
The model generally assigned high confidence scores (e.g., $\geq$ 8 out of 10) to LLM-generated style descriptions and comparisons (where 10 = a perfectly accurate style description of the writing sample, or comparison between two styles).
While adequate for a design probe, practitioners who deploy similar systems for production should rigorously validate LLM output quality~\cite{kim2023evallm}.

While most of our prompts are straightforward,
computing style updates requires additional logic (\autoref{fig:style-update}). 
We first pass the \textit{current style} description, user's \textit{likes \& dislikes} summaries, and \textit{full text document} to the backend. 
The LLM produces a new style description based on the provided text and likes \& dislikes, formatted as HTML (\autoref{fig:edit-style}), along with a short style summary (\autoref{fig:teaser}). 
Next, we ask the LLM to write a comparison of the new and old styles, including a difference rating to determine whether a style update is necessary (\autoref{fig:history}). 
If the rating is greater than a threshold (currently set at 3 out of 10 based on pilot studies), we return the \textit{new style} and \textit{comparison} as outputs to the frontend. 
Otherwise, the system outputs a ``no style update needed'' message. 

\subsection{Design Limitations}
Our current implementation only stores one global style and context, and simplifies style to five dimensions. This may not capture all aspects of the desired writing style. 
However, users are not limited to this (initial) setup, as they can directly edit the system's style and add explicit likes \& dislikes to better guide text generations. 

\section{User Study}\label{sec:study}
We conducted a two-part study with 18 participants, using GhostWriter as a probe to examine how our designed experience shapes user behaviors with and reactions to AI in \textit{editing} and \textit{creative writing} tasks. 
These two tasks were informed by background research and discussions with collaborators about common writing scenarios (1) where personalization is important and (2) that are relevant to a broad range of stakeholders~\cite{clark2018creative,singh2022hide,mirowski2023co,raheja2023coedit,yuan2022wordcraft}.
We also chose these tasks to allow users to (potentially) explore different features in GhostWriter.

Our user study aims to address the following research questions:
\begin{itemize}
    \item \textbf{RQ1:} How does having the ability to affect style and context help users achieve their writing goals?
    \item \textbf{RQ2:} How do users react to the different ways to craft personalized content in GhostWriter?
    \item \textbf{RQ3:} What new challenges emerge for users from interacting with GhostWriter?
    \item \textbf{RQ4:} How do users perceive the relationship between writers and AI?
\end{itemize}

\subsection{Participants}
We recruited 18 participants (16 women, 2 men) via mailing list at a large technology company. 
Our selection process was self-selective and targeted individuals whose professions involve writing in significant ways. 
This led to 7 content designers, 6 UX researchers, 4 communications managers, and 1 executive assistant (\autoref{tab:participants}). All participants were based in the United States. 

\subsection{Procedure}
Participants interacted with GhostWriter during two 1-hour sessions, each focused on a distinct writing task.
Session 1 was designed to familiarize participants with the system, and we erred on the side of building expertise with a shorter \textit{editing} task, before diving deeper into a longer \textit{creative writing} task in Session 2.
We understand the possibility of learning effects between sessions, but as a design probe looking into mostly qualitative insights, we think the benefits of a gradual introduction to GhostWriter outweigh any negative ordering effects.

Sessions took place online using Microsoft Teams where we asked participants to think aloud. We also recorded their screens and logged system events.
Participants were compensated after study completion with a \$100 Amazon gift card. 
This study was approved by our company's Institutional Review Board.

\subsubsection{Pre-study Survey}
Prior to Session 1, participants completed a survey about their experience with generative AI (GenAI) systems.
15 of the 18 participants reported working with GenAI systems for less than one year (one had never interacted with them). 
Five participants reported interacting with such systems multiple times a week, another five reported multiple interactions per month, and the remainder reported less frequent interactions. 

Six participants indicated that they generally write multi-sentence prompts, with the rest writing simpler prompts. 
10 out of 18 participants noted that they do not attach contextual documents when prompting GenAI systems.

\begin{table}
\small
\centering
\caption{Overview of user study participants.}
\label{tab:participants}
\begin{tabularx}{\linewidth}{p{.1\linewidth}p{.6\linewidth}p{.25\linewidth}}
\textbf{ID} & \textbf{Role}                 & \textbf{Gender} \\ \midrule
P01         & Senior content designer       & Female          \\
P02         & Senior content designer       & Male            \\
P03         & Content designer              & Female          \\
P04         & UX researcher II              & Female          \\
P05         & Content designer II           & Female          \\
P06         & Executive assistant           & Female          \\
P07         & Senior communications manager & Female          \\
P08         & Senior UX researcher          & Female          \\
P09         & Senior UX researcher          & Female          \\
P10         & UX researcher                 & Female          \\
P11         & Senior communications manager & Female          \\
P12         & Senior content designer       & Female          \\
P13         & Senior communications manager & Female          \\
P14         & Senior UX researcher          & Male            \\
P15         & Senior content designer       & Female          \\
P16         & UX researcher II              & Female          \\
P17         & Senior communications manager & Female          \\
P18         & Senior content designer       & Female
\\
\end{tabularx}
\end{table}

\subsubsection{Tutorial \& Practice}
Participants started Session 1 by watching a 5-minute tutorial. 
Afterwards, participants had 10-15 minutes to familiarize themselves with GhostWriter. 

\subsubsection{Task 1: Professional Editing}
The main task in Session 1 was to \textit{edit} and refine a document to fit a particular writing style based on these instructions:

\begin{sprompt}{}
\small
     Imagine you are a freelance content writer working with a new client. The client runs a travel blog, \textit{EpicWanderlust}, and wants you to help write a new post about Seattle. You are given a draft with some basic ideas, but the client feels it does not fit the ``style'' of their blog. Here is where you come in: Your job is to polish the draft (with GhostWriter's help!) so that it feels more cohesive and consistent with the other posts on \textit{EpicWanderlust}.
     
\vspace{0.25em}

    \textbf{One additional request:} the client wants you to help them reach younger audiences by tweaking the post's writing style.
    \vspace{0.25em}
\end{sprompt}

\noindent We included an example blog post from \textit{EpicWanderlust} to help participants get started and extract the desired writing style,
along with some default context on Seattle. 
Participants were given $\sim$30 minutes to complete this task.

\subsubsection{Task 2: Creative Writing}
The main task in Session 2 was to \textit{generate} a document following a writing style chosen by the participant. 
We asked participants to bring a short writing sample containing a style they wanted to emulate (written by them or someone else) and then write a story based on one of these prompts:

\begin{sprompt}{}
\small
\vspace{-0.25em}
     \textbf{Story 1:} ``Write a short story about an intern at a tech company having an adventure on an alien planet.''
        \vspace{0.25em}
        
        \textbf{Story 2:} ``Write a short story about a group of friends who find themselves trapped in a haunted mansion.''
        \vspace{0.25em}
        
        \textbf{Story 3:} ``Write a short story about a musician who finds a magical device that can control the weather.''
    \vspace{0.25em}
\end{sprompt}

\noindent Participants chose which prompt to work on, and in each case, we provided some default context about a possible setting and characters to build upon. 
Participants were given $\sim$45 minutes to complete this more open-ended, creative task.

\subsubsection{Post-task Survey}
After each session, participants completed the same, $\sim$10 minute post-task survey. 
The survey started with Likert scale questions (where 1: \textit{strongly disagree} - 5: \textit{strongly agree}) about user satisfaction, ease of use, perceived agency, and output ownership, as well as system trust and understanding.
Then, participants filled out open-ended questions about their overall experience with GhostWriter, likes, and dislikes. 
After completing both sessions, we asked participants how they perceived their relationship with the AI system.
All survey questions are included as Supplementary Material.

\subsection{Data Analysis}
Our approach of surfacing insights through participant task interactions and post-task feedback is well-aligned with established design probe methodology, which prioritizes collecting reflections and reactions from users~\cite{hohman2019gamut,boehner2007hci}.

We adopted a mixed-methods approach for data analysis. 
First, we quantitatively analyzed our event logs to examine how participants interacted with GhostWriter, focusing on event counts (e.g., likes, style updates) and common interaction sequences.
We then performed a thematic analysis by coding recurring patterns in participants' inline and ``apply'' prompts. 
Next, we computed metrics based on post-task Likert scale responses.
Finally, we synthesized themes from the qualitative survey data and think-aloud interview transcripts to collect participant impressions about their experience with our AI-powered writing assistant. 

Thematic analyses were performed inductively, with two authors independently coding a subset of the data to establish initial themes.
A shared codebook was developed through iterative discussion and refinement, which was used to code the remaining data.

\section{Results}\label{sec:results}
All participants (\textbf{P01-P18}) successfully completed both writing tasks. 
The \textit{Editing} task took an average of 28.7 minutes to complete, while the \textit{Creative Writing} task took 46.7 minutes.

\subsubsection*{Overall Impressions}
\begin{figure*}
    \centering
    \includegraphics[width=\linewidth]{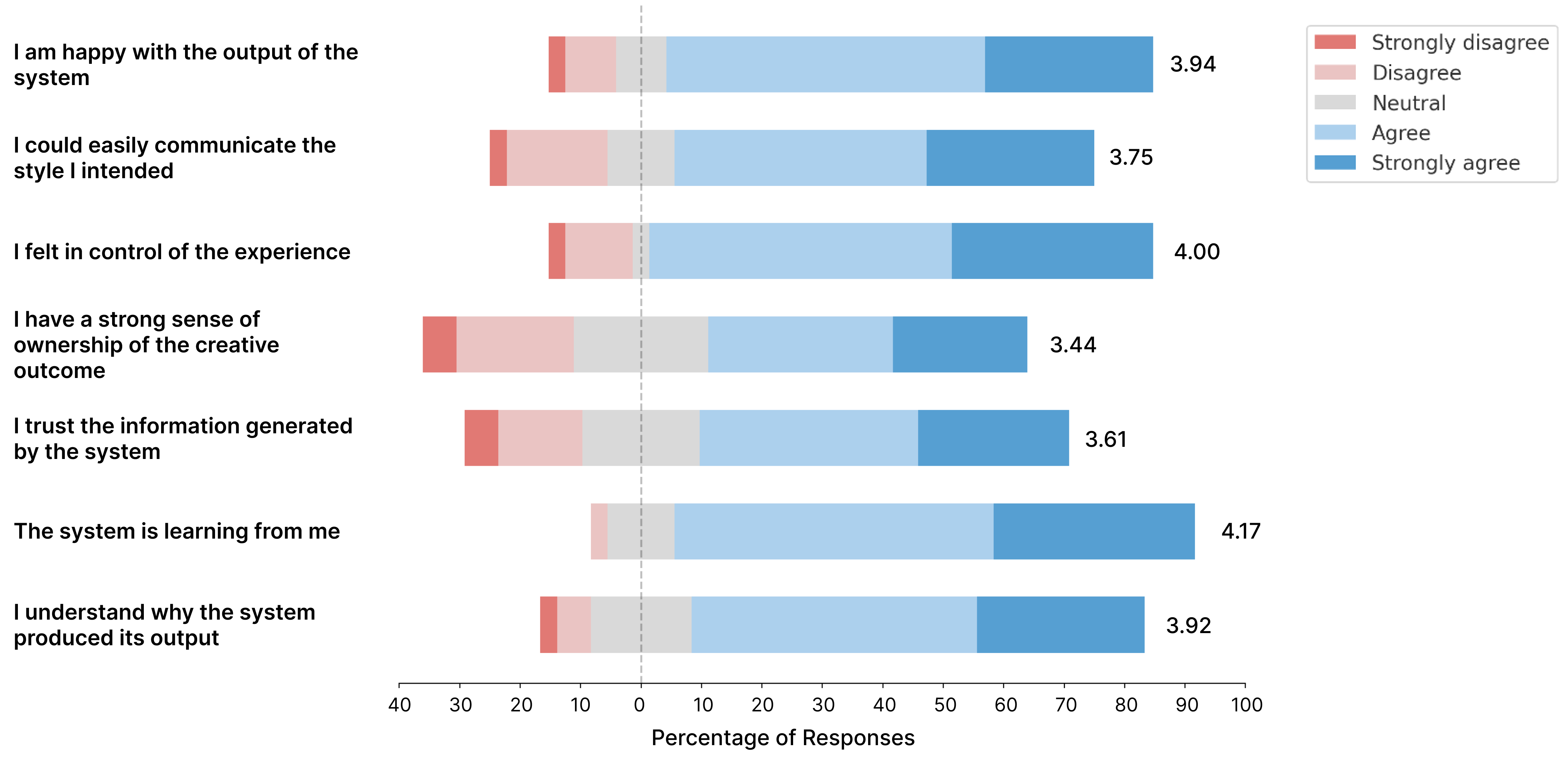}
    \caption{Aggregated participant responses to survey questions regarding system usability and satisfaction after completing each user study task with GhostWriter. Each question was scored on a 5-point Likert scale (1: \textit{strongly disagree} - 5: \textit{strongly agree}). Each horizontal bar contains 36 responses (2 per participant), with average values displayed on the right.}
    \Description{Horizontal bar chart showing aggregated participant Likert survey question responses regarding system usability and satisfaction.}
    \label{fig:survey}
\end{figure*}

Participants gave positive responses about their experience with GhostWriter (\autoref{fig:survey}). 
We visualize aggregated responses from both tasks, since our goal was to collect impressions about GhostWriter as a general writing tool.\footnote{We also did not observe significant differences in scores after each task.}
The statements, ``I trust the information generated by the system'' (mean: 3.61), along with ``I have a strong sense of ownership of the creative outcome'' (mean: 3.44), yielded the most variation and lowest mean ratings. 
In each case, however, 61.1\% and 52.7\% of responses were still ``agree'' or ``strongly agree,'' respectively.

``I felt in control of the experience'' (mean: 4.00) and ``The system is learning from me'' (mean: 4.17) received the highest mean scores for both tasks, reflecting positively on our goals of \textbf{personalization} and \textbf{agency}. 
83.3\% of responses were ``agree'' or ``strongly agree'' with the former statement, and 86.0\% showed agreement with the latter.
Ratings indicated that most participants were able to communicate their intended style (mean: 3.75), were satisfied with GhostWriter's customized text generations (mean: 3.94), and understood the system's behavior (mean: 3.92).

\subsection{RQ1: How does having the ability to affect style and context help users achieve their writing goals?}\label{sec:results_rq1}

Participants iteratively crafted their desired style through implicit or explicit methods -- showing how \textbf{agency over style manifested in different ways} throughout GhostWriter. 
There was an average of 6.71 style updates per task: 3.39 were automatic, 1.87 were direct edits, and 1.45 were manually requested.
During each task, users also added an average of 3.77 likes and 2.81 dislikes.
This suggests participants were not just passive recipients of stylistic suggestions, but engaged in intentional, often mixed-mode refinement to guide GhostWriter toward their individual goals. 
On average, participants viewed the style history page 0.97 times per task, and the likes \& dislikes page 1.97 times. 
In contrast, they viewed the main style page an average of 4.99 times, showing a preference for in-the-moment control over retrospective style comparisons.

\subsubsection{Editing vs. Creative Writing}
Participants viewed the context page on average 0.86 times during \textit{Editing}, compared to 4.41 times in \textit{Creative Writing}.
This shift -- illustrating how user agency manifests through \textbf{adaptive context-seeking behavior} -- is reasonable given the tasks' differing natures; during \textit{Creative Writing}, participants frequently referenced context information for composing text, while during \textit{Editing}, they focused on refining the text.

On average, participants used the rewrite feature 3.52 times per task. 
The continue feature was used less in \textit{Editing} (1.14 times) compared to \textit{Creative Writing} (2.59 times).
Similarly, for \textit{Editing}, participants used an average of 1.14 inline prompts, while for \textit{Creative Writing}, they used an average of 5.18. 
The latter's focus on content creation likely explains the higher usage of features that support \textbf{idea generation and progression}.
However, participants used an average of 4.57 ``apply'' prompts during \textit{Editing}, compared to 3.47 in \textit{Creative Writing} -- 
reflecting a preference for more \textbf{targeted forms of style control} during the former task.

\subsubsection{Intents By Prompt Type}\label{sec:results_rq1_prompt}
To better understand usage differences between our two LLM prompt features and the intents behind them, we analyzed the 97 inline and 134 ``apply'' prompts composed by participants across both tasks.
Overall, we found that participants used inline prompts for \textbf{brainstorming and creative agency}, and ``apply'' prompts for \textbf{precise style refinement}.

For inline prompts, the most common intent was \textit{adding more content} ($n=42$; 43\%), e.g., ``Elaborate on how John comes across PlayStation while exploring Starfield'' or ``Add a paragraph with additional places to visit.'' 
Participants frequently wanted to \textit{generate full drafts} of documents ($n=30$; 31\%), e.g., ``Write a horror story in the style of Edgar Allan Poe. The story should have a strong plot with a surprise twist'' or ``Write a blog article with an introduction, 4 paragraphs with catchy titles and a summary to convince my friend to visit Seattle.'' Ten inline prompts (10\%) asked for a document \textit{introduction}, and nine (9\%) asked for a \textit{conclusion}.

With ``apply'' prompts, users often wanted to \textit{expand} the selected text ($n=53$; 40\%), e.g., ``Add more details about his fellow researchers; reference a second character named Bob.'' 43 prompts (32\%) aimed to \textit{rewrite} the selected text, e.g., by changing the perspective (``Change to first person''), tone (``Make this more positive and enthusiastic''), or audience (``Rewrite for younger readers''). Participants also used ``apply'' prompts to \textit{condense} text ($n=6$; 4\%), e.g., ``Shorten this poem by half'', or request specific \textit{formatting} ($n=7$; 5\%), e.g., ``Bullet form this paragraph.'' 
In 3 cases (2\%), users wanted to \textit{add transitions} between paragraphs.
3 prompts (2\%) asked for \textit{suggestions or critique}, e.g., ``Is this a well-written sentence?''

\subsubsection{Interaction Patterns Over Time} 
Writing tasks are not monolithic, and people's behaviors (like any story) can shift over the course of a task.
As such, we looked at \textbf{differences in interaction patterns between the first and second half} of each task. 

Participants requested manual style updates more frequently in the first half (1.19 times) vs. the second half of tasks (0.29 times; $W = 10.0, p < .004$).\footnote{We report Wilcoxon rank-sum test statistics and use a Bonferroni correction of $\alpha = \frac{0.05}{13} = .004$, as we perform one test per $n=13$ log event types.}
Similarly, visits to the main style page dropped from an average of 3.32 times in the first half of each task to 1.68 in the second half ($W = 72.5, p < .004$). 
The context page followed a similar trend, with 2.10 visits in the first half compared to 0.71 in the second half ($W = 24.0, p < .004$). 
Conversely, visits to the likes \& dislikes page increased from 0.58 to 1.39 between task halves ($W = 65.0, p = .01$).
These observed behaviors reflect how during early-phase exploration and personalization, users may employ various techniques to shape their desired writing style and context.
However, as writing progresses, they prefer more lightweight methods for finetuning the system's style knowledge.

\begin{figure*}
    \centering
    \includegraphics[width=\linewidth]{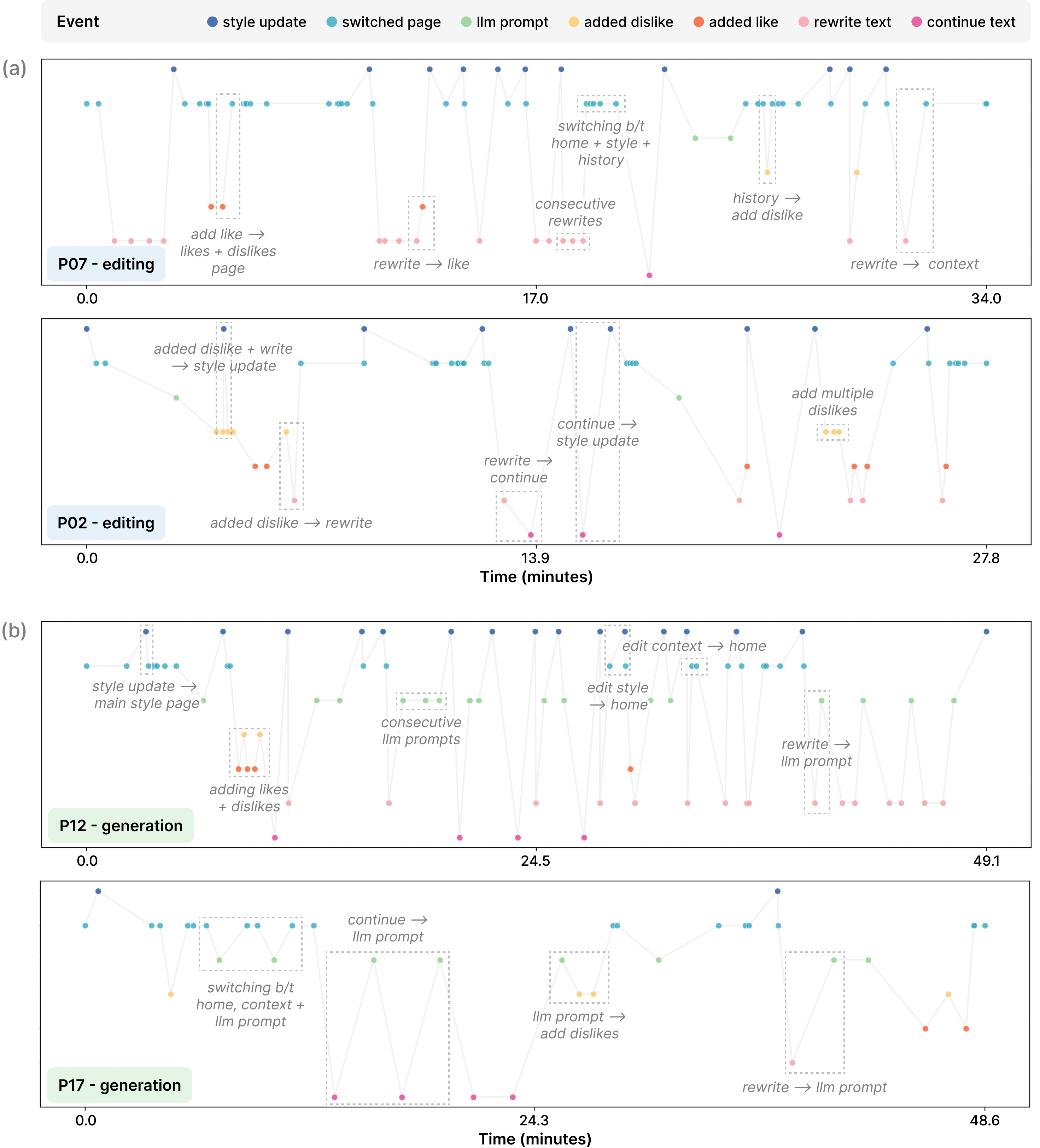}
    \caption{Example participant telemetry timelines from the \textbf{(a)} editing and \textbf{(b)} generation tasks. Key events are plotted in different horizontal lanes to avoid visual overlap, and time is plotted along the x-axis. Notable interaction patterns are highlighted with the gray dashed boxes. 
    }
    \Description{Example participant telemetry timelines from the (a) editing and (b) generation tasks. Key events are plotted with unique colors and in different horizontal lanes to avoid visual overlap, and time is plotted along the x-axis.}
    \label{fig:timeline}
\end{figure*}

\subsubsection{Responding to System Feedback}
We plotted telemetry timelines to gain additional insights into participant interaction patterns (\autoref{fig:timeline}).  
Across both writing tasks, users viewed the main style page 61 times (36\% of the time) directly after a manual or automatic style update.
After editing the description on the style page, participants returned to the home document 54 times (71\%). 
Similarly, visiting the context page was often followed by navigating to the home (58 times = 55\%) or style page (25 times = 24\%). 
These patterns suggest users' desire to \textbf{examine or reflect} on the system's state when it changes, and eagerness to \textbf{test GhostWriter's knowledge} after revising style and context.
Users also frequently added consecutive likes and dislikes to steer style (160 times = 70\%). 
However, after adding a highlight, participants only viewed the likes \& dislikes page 72 times (17\%), indicating a potentially reduced need or motivation for reflection in these cases.

\subsubsection{Strategies for Personalization}
Our analysis highlights differences in how participants experienced personalization and agency when interacting with GhostWriter. 
For example, P07 and P12 used the rewrite feature more extensively compared to P02 and P17 while writing (\autoref{fig:timeline}). 
After rewriting text, participants also took different next actions, using a ``apply'' prompt 21 times (18\%; e.g., P12, P17), adding a like 19 times (17\%; e.g., P02, P07) or dislike 13 times (11\%), and performing another rewrite 17 times (15\%; e.g., P07, P12). 
Some participants (e.g., P07, P12) liked using two inline or ``apply'' prompts in a row to generate personalized content; the former occurred 28 times (25\%) and the latter 31 times (23\%).  

Users added likes and dislikes at different points, highlighting the value of allowing \textbf{flexible moments for user agency} during writing.
P12 mainly added them at the start of the task, while P02 and P07 did so at both the beginning and end, and P17 spread their likes and dislikes throughout the session.

\subsection{RQ2: How do users react to the different ways to craft personalized content in GhostWriter?}\label{sec:results_rq2}
Several participants described the potential of our system to \textbf{augment human creativity and productivity} through content personalization ($n=8$): \textit{``[I] would definitely use this kind of tool to help me be more productive in a work setting''} (P17). 
P02 added, \textit{``I think [GhostWriter] offers a good starting point for any kind of writing,''} and P07 loved \textit{``seeing how creative it can be and how it boosted my own imagination.''}

\subsubsection{Ease of Use and Style Learning}
Overall, participants viewed GhostWriter as \textbf{easy to use and intuitive} ($n=15$). P04 described our system as \textit{``fun and robust,''} and P14 reported, \textit{``This was an incredibly positive experience. It was easy to build and iterate on the story. The tool has an easy learning curve, and a simple interface.'' }

Participants were particularly excited about the ease of \textbf{style learning} ($n=18$). 
P06 said, \textit{``I was impressed how the system understood and incorporated my style, both from the sample text and [my own] writing.''} 
Similarly, P09 appreciated GhostWriter's \textit{``responsiveness to the style guide''} and P18 was \textit{``pleasantly surprised at how much content I was able to generate in a style I preferred.''} 
P12 mentioned \textit{``[GhostWriter was] helpful for real time feedback on [writing] style''} as well. 

\subsubsection{Multiple Personalization Paths}
Consistent with the results from our timeline analysis (Section~\ref{sec:results_rq1}), participants enjoyed having different ways to interact with style and provide feedback ($n=9$), \textbf{valuing the agency to select between methods} \textit{``that are optimal for different contexts and objectives''} (P09).
7 participants particularly appreciated the likes \& dislikes feature, which was perceived as \textit{``easy to use and effective''} for tailoring LLM outputs (P08). 
Users also frequently filled out the optional feedback field, i.e., specifying why they highlighted a portion of text. 

The context page ($n=7$) and ``apply'' prompt ($n=4$) were other common favorite features. 
P08 said the former \textit{``helped organize my thinking,''} and P10 thought that \textit{``context adds value into the prompt.''}
Participants like P07 were \textit{``fascinated by how [GhostWriter] combines the sample, style, and context to find the right words.''}

\subsection{RQ3: What new challenges emerge for users from interacting with GhostWriter?}\label{sec:results_rq3}
Our study with GhostWriter revealed challenges around designing for personalization and agency that may be relevant to other AI-enhanced writing systems.

\subsubsection{Confusion Over Style and Context Interactions}
Many participants explored personalization through editing both style and context information. 
However, some did not fully grasp the \textbf{role of style vs. context} ($n=4$), which can limit the effectiveness of these personalization controls: \textit{``[I'm] trying to understand how style impacted context when they were conflicting in tone''} (P15), or \textit{``[I'm not] sure if I should put [information] in the context or style''} (P04). 
Others like P16 hypothesized that \textit{``style has more influence than context''} in steering text generations.

Participants also expressed confusion about how GhostWriter behaved after learning a new style. 
To preserve agency, our system never rewrites documents without explicit user action.
However, P14 thought that after a style update, GhostWriter would automatically ensure \textit{``[all] the text gets updated''} as well. 
P03 added, \textit{``Refreshing the style [and] applying it was not very easy to wrap my mind around.''} 
The impact of \textbf{explicitly refreshing style}
was also unclear to some: \textit{``If I hit refresh, is it going to extract the style of this [document] or apply the style I pasted?''} (P15). 

\subsubsection{Need for Contextual Awareness}
Another issue was GhostWriter's occasional \textbf{lack of contextual awareness} ($n=5$) when generating content in the middle of a document or building on existing text (e.g., ``Explain how Riley dealt with the negative impact on his internship'' or ``Finish the story with the characters running outside''). 
P02 noted how GhostWriter \textit{``was unable to continue writing content from the previous sentence. It just wrote more [unrelated] text.''} 
Similarly, P08 hoped inline generations could start \textit{``at a midpoint [vs.] always having to restart.''}
This is a limitation of our current implementation, but it highlights a key consideration for designing effective personalized writing experiences.

\subsubsection{Additional Style and Personalization Controls}
Participants wanted GhostWriter to support a richer personalization infrastructure, e.g., by \textbf{storing multiple styles and offering preset style templates} ($n=8$).
Particularly for users working on different writing tasks and crafting diverse documents, it could be helpful to \textit{``apply context/style to individual documents''} (P05) or \textit{``allow different styles for different parts of the document''} (P08). 
P10 suggested having a style ``library'' to facilitate applying different styles, similar to document templates in Microsoft Word or Google Docs. 

Six participants requested \textbf{alternative forms of style expression}.
Some mentioned that GhostWriter's open-ended way of specifying style through natural language could be a drawback: \textit{``it was difficult to articulate a style that I wanted to replicate''} (P09).
P04 wanted to \textit{``make style more structured,''} while P08 and P10 proposed having \textit{``suggestions on styles (drop-downs of various options to spark individual creativity).''} 
P15 suggested \textit{``quantifying''} style using \textit{``variables that can be defined or fed [like] audience or venue.''} 
Five participants wanted to \textbf{validate style accuracy} as well. 
As P15 expressed, it may be helpful to include \textit{``some kind of a score---how close are you with the style intended.''}

Participants also wanted \textbf{more fine-grained options} for asserting agency over LLM outputs ($n=11$). P06 said, \textit{``I wish I could've accepted only parts of the text it generated [and] provided feedback so it could learn what I liked and didn't.''} 

\subsection{RQ4: How do users perceive the relationship between writers and AI?}\label{sec:results_rq4}
After both sessions, we asked participants how they view the role of AI in a system like GhostWriter. 
These perspectives appear to be closely related to how users engaged with our probe while writing. 

Eight participants saw AI as a \textbf{tool} due to observing GhostWriter's reliance on human prompts and the lack of collaboration in generating outputs: \textit{``It's very powerful in generating well-written text. But it [needed] instructions to do that''} (P11). 
P08 added, \textit{``it would become more of a collaborator''} if they \textit{``spent more time adding my own text and getting to the point where we were working together.''} 
P16 explained, \textit{``Collaborator feels too strong. [The system] feels more like a sounding board. But I don't think it actually understands my ideas,''} emphasizing the human-AI communication gap and a more utilitarian relationship where the user maintains creative agency.

Conversely, 6 participants, who iterated with the system or explored multiple directions, framed it as a \textbf{collaborator}: \textit{``It helps you get started and then you can react to it''} (P12). 
Similarly, P07 said systems like GhostWriter could help generate ideas and \textit{``boost creative thought if you have writer's block.''}
P15 saw AI as a \textit{``second [pair] of eyes, like [how] you run presentations by your colleague''} but would view it as more of a \textit{``potential collaborator''} if there was more back and forth between writer and system -- suggesting that dialogue is a key factor in perceived co-creation.

Four participants regarded AI as \textbf{both} a tool and collaborator, and 3 thought AI could take an \textbf{advisor} role as well, e.g., if it gave \textit{``feedback or ideas [to] improve''} like pointing out \textit{``plot holes or inconsistencies'}' (P04). 
This aligns with participants' use of GhostWriter to ask for writing critiques and suggestions (Section~\ref{sec:results_rq1}).

\section{Design Recommendations \& Discussion}\label{sec:discussion}
We present design recommendations for creating similar AI-supported writing experiences, shaped by our findings and overarching themes of \textbf{personalization} (Section~\ref{sec:guidelines_personalization}) and \textbf{agency} (Section~\ref{sec:guidelines_agency}). 
We also share observations about how these experiences can better support reflection (Section~\ref{sec:reflection}), and insights on how people view the role of AI and ownership in the context of writing (Section~\ref{sec:ownership}).

\subsection{Designing for Personalization on People's Terms}\label{sec:guidelines_personalization}
\subsubsection{Providing multiple paths to personalization is important.} 
One of GhostWriter's perceived strengths was supporting different ways to personalize a desired target style (Section~\ref{sec:results_rq2}).
As users differ in their writing workflows and preferences, it makes sense to offer agency in how they can interact with AI writing systems by including both \textbf{implicit} and \textbf{explicit} paths~\cite{lee2024design}. 

Our work surfaces how this flexibility is especially beneficial because one's preferred mode of controlling style may change throughout the writing process. 
For example, manual style and context edits generally decreased throughout each task, suggesting these forms of style tweaking may be more useful earlier in the writing process, while users are exploring and defining their creative direction (Section~\ref{sec:results_rq1}). 
On the other hand, users highlighted more likes \& dislikes as writing progressed, pointing to the importance of \textbf{temporal personalization} -- adapting support based on writing phase -- in future AI-infused systems.

\subsubsection{Explicit teaching moments can be worth the effort.} 
Users regularly engaged with the likes \& dislikes feature (\autoref{fig:timeline}), which was unexpected, as annotations require extra effort during writing.  
Several noted that this was their favorite feature, illustrating the value of offering \textbf{explicit feedback opportunities} in AI-augmented systems. 
Participants even requested additional opportunities to provide feedback, i.e., \textit{``directly on the text [GhostWriter] generated''} (P08) and \textit{``at all input levels''} (P06). 
This finding aligns with the guideline ``Encourage granular feedback'' from~\citet{amershi2019guidelines} and principles for leveraging smaller ``units of information''~\cite{rachatasumrit2021forsense}, suggesting that such mechanisms are key toward the goal of personalization.
Moreover, our observations indicate that explicit feedback empowers users to assert agency in identifying and underscoring personal style dimensions that the system's extraction process might miss (e.g., audience or formatting).

\subsubsection{Substance and style are important. So is format.} 
Although our system analyzes sentence and paragraph structure when updating style, it currently does not incorporate other aspects of formatting. 
However, many participants viewed document \textbf{formatting as an integral part of style}, wishing to use GhostWriter's features to personalize and affect the format of generated text (Section~\ref{sec:results_rq1}). 

Formatting was also used as a non-literal part of participants' (teaching) language when explicitly editing the system's style (\autoref{fig:edit-style}b), e.g., bolding words to underscore their importance. 
This type of weight specification is consistent with observations by~\citet{Ng2020UnderstandingAS} on teaching languages and presents a promising avenue for giving users richer expressive control over personalizing style.
We see connections to work such as Textoshop~\cite{masson2024textoshop}, which use drawing software-inspired interactions to stylize writing.

\subsubsection{Having one style is good. Having many is better.} 
Several participants thought GhostWriter could be even more powerful if it allowed them to define and opportunistically select \textbf{different styles and contexts}, rather than having one global profile for all documents (Section~\ref{sec:results_rq3}). 
Going from having one style and context to many is a trend in the emerging capabilities of independent agents like OpenAI's custom GPTs.\footnote{\url{https://openai.com/blog/introducing-gpts}} 
Having the ability to invite these different styles and contexts into the same document is the natural next step to satisfy users, enabling AI-powered systems to support and personalize a wider range of writing tasks~\cite{olatunjiinteractive,lee2024design}.

\subsection{Designing for Layered and Contextual Agency}\label{sec:guidelines_agency}
Preserving user agency was top of mind when creating GhostWriter. 
However, we observed some trade-offs when designing for agency, which warrant further investigation.

\subsubsection{Natural language as an interface can enhance agency. It can also add unwanted effort.} 
We initially hypothesized that open-ended style expression would provide users with empowering flexibility. 
However, some participants found it challenging to articulate style through free-form natural language, and desired a strongly structured format such as dropdown selectors---even at the cost of reduced agency (Section~\ref{sec:results_rq3}). 
From our results, we see room to investigate \textbf{different style specification languages}, which may depend on a user's role and background, and the particular writing scenario.
One possibility is to support mixed-initiative style configuration -- where natural language can be supplemented with visual scaffolds, example-based style selection, or other interactive controls -- to reduce cognitive effort without compromising expressivity~\cite{clark2018creative,singh2022hide}.

\subsubsection{Consider expectations regarding system behavior after learning a new style.} 
Our decision to prevent automatic document rewrites after style changes contradicted some participants' expectations (Section~\ref{sec:results_rq3}), underscoring an \textbf{agency / transparency vs. responsiveness trade-off}. 
On one hand, instantly refreshing documents after style updates could reduce users' perceived sense of control and awareness when using GhostWriter, decreasing system trust~\cite{richard2024agency}. 
On the other hand, people want immediacy in their actions and to avoid the repetitive steps of style change and application.
We advocate for designs that balance agency and awareness during AI-assisted writing, e.g., by previewing changes to users and working to align system behavior to their mental models.

\subsubsection{Agency is desired at different levels.} 
Our work focuses on providing agency in style and context definition.
After experiencing this agency, participants requested similar control over \textbf{refining LLM generated outputs} (Section~\ref{sec:results_rq3}), which is consistent with~\citet{yuan2022wordcraft,ippolito2022creative,gmeiner2023dimensions}.
Our observations and the intrinsic, iterative nature of writing underscore the importance of providing users with tools to confidently steer LLM outputs at multiple levels of text generation and style crafting: draft, intermediate, and final~\cite{gmeiner2023dimensions,du2022read,lee2024design,gero2022design}. 
Users may also require different kinds of agency across different tasks (e.g., fine-grained control for editing vs. high-level ideation support for creative writing -- Section~\ref{sec:results_rq1}).

\subsection{Supporting Reflection During AI-Mediated Writing}\label{sec:reflection}
To create opportunities for reflection (see \textbf{DP4}, Section~\ref{sec:system}), we included features such as the main style page, likes \& dislikes page, and style history page where users can explore learned styles. 
The style page was frequently viewed by participants (\autoref{fig:timeline}), who enjoyed examining its content and seeing how (and how well) GhostWriter interpreted their style (Section~\ref{sec:results_rq2}).
However, our other reflection features were less utilized, potentially due to the cost of information processing (Section~\ref{sec:results_rq1}). 
As P15 said, \textit{``These pages [aren't] telling me much, or at least I didn't have the patience to go through all of it.''} 
P14 shared, \textit{``It would be cool if you could toggle to see the style, history, or likes in the right nav so you don't have to go between the home page and other options,''} suggesting ways to more seamlessly integrate \textbf{lightweight reflection mechanisms} into the natural flow of writing~\cite{teufelberger2024llm}.

Overall, our work highlights how reflection can \textbf{help people understand and act on AI responses}~\cite{liu2023incorporating,mahlow2023writing,xi2025omnithink,amershi2014power}, pointing to important design opportunities. 
For example, given the literature on writing as a technology for thinking (e.g.,~\citet{ong1992writing}), it seems worthwhile to provide interactions for writers to reflect on their style and how it changes across a document or over time. 
Additionally, when multiple styles are learned  (Section~\ref{sec:guidelines_personalization}), future work can explore how writers decide what constitutes a desirable or undesirable style in a particular context. 
Systems such as TextFocals~\cite{kim2024towards} also reveal opportunities for using adaptive AI-generated views to encourage
reflection and self-driven revision of writing without directly generating text with LLMs.

\subsection{Considering the Role of AI and Ownership in AI-Infused Systems}\label{sec:ownership}
Our study allowed us to investigate the evolving relationship between writers and AI, focusing on the connection between \textbf{perceived ownership} and \textbf{process agency}.

\subsubsection{The role of AI in collaborative writing}
For any collaborative human-AI task, it is crucial to \textbf{study perceptions about AI} and design with this information in mind.
Given participants' diverse perspectives on the relationship between writers and AI (Section~\ref{sec:results_rq4}), we encourage future work to look at shaping AI-mediated experiences to help AI fulfill different roles~\cite{olatunjiinteractive,lee2024design,wan2024felt}. 
For instance, how can we transform LLM-powered writing systems to help AI serve more as collaborators rather than tools?

Works such as~\citet{guo2024pen,biermann2022tool,chakrabarty2024creativity} suggest that \textbf{the role of AI can fluctuate} throughout the writing process and depends on an author's values regarding which parts of their writing they wish to maintain control over (e.g., craftsmanship and authenticity). 
We observed a similar phenomenon with GhostWriter, where many users switched preferred modes of style personalization over the course of writing, and between different tasks (Section~\ref{sec:results_rq1}).

Exploring this dynamic interplay between user values, agency, and perceptions of AI is critical to supplement and extend existing research on designing AI-enhanced productivity and creativity support tools (e.g.,~\citet{chung2022artist,lawton2023tool}).
This fluctuation also suggests the need for systems to support role fluidity -- allowing users to move between different types of AI assistance (e.g., exploratory vs. prescriptive) as their goals and writing evolve.

\subsubsection{Agency over the writing process influences one's sense of outcome ownership, but is not sufficient to define it.} 
With GhostWriter, we explored whether preserving agency could increase users' perceived ownership over the results of collaborative AI writing. 
In some cases, this was true: \textit{``I do [feel ownership] because there were so many different things I could inform to make [it] my own''} (P18), or \textit{``I didn't actually create''} the document, but \textit{``I created the prompt. Yeah, I'm putting my name on it''} (P01). 
Others were hesitant to claim ownership over the generated text: \textit{``I architected it, but I didn't build it''} (P09), or \textit{``There's something about not feeling like the owner when using a product like this - more like a partner I should split the by-line with''} (P13). 
P04 shared that they would only feel ownership \textit{``if I weren't using the tool.''}

The \textbf{tension between ownership and agency} raises questions about how ownership should be perceived and (possibly) redefined in the age of generative AI~\cite{wasi2024llms}.
\citet{draxler2024ai} frames this tension as the ``AI ghostwriter effect,'' where users resist labeling AI-assisted writing as AI-authored, but also do not consider themselves the owner of this work.
However, the authors find that increasing user control over co-written text bolsters perceived ownership, mirroring the sentiments of participants like P18 when using GhostWriter's various style controls.
Similarly,~\citet{joshi2024writing} shows that crafting longer, detailed prompts enhances user ownership, but this requires more metacognitive effort~\cite{zamfirescu2023johnny}.
Counterquill~\cite{ding2024counterquill} demonstrates that breaking down the human-AI writing process into three distinct phases --- learning, brainstorming, and co-writing --- can also increase writers' perceived ownership.

These findings, along with those from~\citet{palani2024evolving}, suggest that to preserve ownership, AI-powered systems should assign writers a \textbf{key orchestrator role} throughout the writing process. 
This role can empower writers to regulate and navigate the shifting boundaries of initiative and responsibility in human-AI co-creation.
GhostWriter takes initial steps in this direction, but our results highlight opportunities to build on this foundation, encouraging the development of systems that more flexibly support user agency, authorship, and creative control in AI-infused writing.

\section{Limitations \& Considerations}\label{sec:future}
Our findings provide encouraging signals on how design can enhance alignment between people and AI writing assistance, while supporting user agency throughout the experience. 
By design, design probes are not intended to serve as comparisons to a baseline; instead, they function as instruments to explore ideas and formulate new questions during early stages of research. 
We consider our work a starting point for comparative studies and future research to build upon.
More work is also needed to validate our results in other types of writing activities and over longer-term use. 

The focus on UX-centric professionals from the United States may restrict the generalizability of our findings to more diverse populations.
Since all participants were employees of a large technology company, they may have more exposure to generative AI than the average person. 
Our participant pool also exhibited a gender imbalance, possibly reflecting biases in certain industry roles.
However, irrespective of role and self-reported background, we did not observe noticeable differences in participants' abilities to access GhostWriter's features. 

Designing personalized systems of any kind requires careful consideration of \textbf{data privacy trade-offs}. 
While personalization offers numerous benefits, e.g., learning and applying custom writing styles, it also poses risks. 
After engaging with GhostWriter, P14 asked, \textit{``If people put sensitive data in here, is it safe?''} 
Many users may not even be aware of the privacy risks associated with such data-driven systems. 
To mitigate these challenges, we advocate for a design approach rooted in transparency and user consent. 
By providing users with clear communication about how their information is used and stored, as well as global controls to turn on/off data collection~\cite{amershi2019guidelines,jorke2023pearl,richard2024agency}, we can build human-centered, personalized experiences that protect user privacy and trust.

\balance
\section{Conclusion}\label{sec:conclusion}
In this work, we explore how to design AI-assisted writing experiences that allow control over \textbf{personalization} and champion user \textbf{agency}. 
Following a set of design goals and principles, we created GhostWriter, an AI-infused editor, and used it as a design probe to study the potential of large language models in crafting personalized writing experiences through style and context.

A study in which participants used GhostWriter on two writing tasks revealed that they valued the tool (arguably compared to previous writing experiences and tools from their daily lives) and perceived it as allowing them to exert rich agency and align AI writing to their goals. 
Participant feedback illustrated the benefits of offering both implicit and explicit mechanisms to teach the system about one's writing preferences -- particularly as these preferences may evolve throughout the writing process.
Guided by these findings, we present design lessons to help others build the new generation of collaborative AI writing systems.

We hope that our work inspires others looking to harness generative AI to enhance and complement human capabilities, providing a reference for exploring the challenges and opportunities that arise when designing and using these emerging technologies.

\bibliographystyle{ACM-Reference-Format}
\bibliography{ref}

\end{document}